\documentclass{article}

\usepackage{microtype}
\usepackage{graphicx}
\usepackage{subfigure}
\usepackage{booktabs} 
\usepackage{amsmath} 
\usepackage[semicolon]{natbib}
\usepackage[frozencache]{minted}
\usepackage{xspace}
\usepackage{stmaryrd}
\usepackage{amssymb}
\PassOptionsToPackage{hyphens}{url}\usepackage{hyperref} 

\usepackage[accepted]{icml2019}

\def\mPyPl{\textbf{mPyPl}\xspace}
\def\kw#1{\textbf{#1}}
\def\str#1{\llbracket #1\rrbracket}
\def\md#1{\mathcal{M}_{#1}}
\def\mdetc#1{\md{#1,\dots}}

\setminted{fontsize=\fontsize{8pt}{8pt}} 
\def\tinyfont{\fontsize{7pt}{7pt}}

\begin{document}

\twocolumn[
\title{\mPyPl: Python Monadic Pipeline Library for Complex Functional Data Processing}
\author{Dmitry Soshnikov, Microsoft, dmitryso@microsoft.com \\ Yana Valieva, Microsoft, yana.valieva@microsoft.com}
\date{\vspace{-0.2in}}
\maketitle

\begin{abstract}
In this paper, we present a new Python library called \mPyPl, which is intended to simplify complex data processing tasks using functional approach. This library defines operations on lazy data streams of named dictionaries represented as generators (so-called \textbf{multi-field datastreams}), and allows enriching those data streams with more 'fields' in the process of data preparation and feature extraction. Thus, most data preparation tasks can be expressed in the form of neat linear 'pipeline', similar in syntax to UNIX pipes, or \texttt{|>} functional composition operator in F\#.

We define basic operations on multi-field data streams, which resemble classical monadic operations, and show similarity of the proposed approach to monads in functional programming. We also show how the library was used in complex deep learning tasks of event detection in video, and discuss different evaluation strategies that allow for different compromises in terms of memory and performance. 
    
\mPyPl library with some documentation and intro videos are available at \url{http://github.com/shwars/mPyPl}{} and \url{http://shwars.github.io/mPyPl}.
\\

\end{abstract}
\vspace{0.4in}
]
\clearpage

\section{Introduction}

It is estimated that in the work of a data scientist, about 80\% is spent on data preparation tasks \citep{CrowdFlower2016}. That is why in any data science language and/or framework many useful tools and libraries exist that simplify the task of data manipulation.

In many practical machine learning tasks we start with the ingestion of the original data, and then enrich that data with some additional features, computing them from the original source. One common concept that allows doing so is called \textbf{dataframe}, which is essentially analoguous to a spreadsheet, with some additional computed columns. In Python, dataframes are well supported by Pandas \citep{mckinney2010}. Also, Pandas supports reading the data source from many popular formats, such as csv, Excel, etc.

However, when handling large volumes of unstructured or semi-structured data, for example in video processing tasks, this approach becomes problematic. First, dataframe typically needs to fit in memory of one machine, which cannot be the case in video or even large scale image processing. Thus, it is essential to be able to load the data on demand, in the form of batches, yet it is desirable to be able to express the data pipeline steps in a clean and concise way as operations on the whole dataset.

The way those tasks are currently handled in deep learning practice is by creating special classes/adapters for dynamic loading of data. For example in Keras \citep{chollet2015}, there is a concept of \kw{ImageDataGenerator}, that allows doing some data augmentation, and can take images on the fly from a specified directory. A sample (slightly simplified) code to use \kw{ImageDataGenerator} is presented below:

\begin{minted}{python}
datagen = ImageDataGenerator(rescale=1./255, ...)

generator = datagen.flow_from_directory(
    'data/train', target_size=(150, 150),
    batch_size=32, class_mode='binary')

model.fit_generator(generator, ...)
\end{minted}

If we want to add some custom augmentation code, or create generator for handling video streams, or produce pairs of images for training Siamese networks --- we would need to create our own custom data generator class (maybe by subclassing existing one). This would immediately require some familiarity with Keras internals.

The fact that we can modify the functionality of existing data generator only by subclassing is inherent to object-oriented design of the Keras library. This can also be said about many other Python libraries and tools. However, we strongly believe that \textbf{for creating data preparation pipelines a functional approach should be used}, when required functionality is achieved not by subclassing, but by composing together simpler processing steps in a functional way. This almost never requires writing custom processing classes or functions, but rather calling existing combinators with some small pieces of code written as lambda-expressions.

To simplify writing such data preparation pipelines, we have created a Python library called \mPyPl, based on functional design and a number of core principles that we will describe in the following sections.

\section{mPyPl Library}

Essentially, \mPyPl is based on three main ideas:
\begin{itemize}
    \item using functional programming techniques and lazy pipelines based on \textbf{Pipe} package \citep{Pallard2016}
    \item using generators that produce streams of named dictionaries (instead of atomic values), which 'flow' through the pipeline (we call them \textbf{multi-field datastreams})
    \item using a small number of basic operations (the most important one being \kw{apply}) that operate on those fields, as well as a number of pre-defined data producers and sinks, hiding the internal implementation complexity
\end{itemize}
    
The main advantage of this approach is the ability to create pipelines that combine several streams of data together.

\subsection{Pipe package}
\label{sec_pipe}
Pipe package allows writing chained computations in the infix form using \verb!|! pipeline operator, for example:

\begin{minted}{python}
    from pipe import *
    res = (range(10) 
          | where (lambda x: x%2==0) 
          | select (lambda x: x*x) 
          | add )
\end{minted}

Here \verb!select! is another name for a \kw{map} function, and \verb!where! stands for \kw{filter}, so this piece of code computes sum of all even numbers smaller than 10.

Such a way of expressing computations looks very powerful, for example, we can easily express the logic of preparing a bunch of images for neural network training, together with some augmentation\footnote{We do not consider augmentation algorithm here, because it is not important for our discussion, we just assume \texttt{augment} function does the job, and \texttt{resize} function does additional resizing.}, as shown below:

\begin{minted}{python}
import mPyPl as mp
images = (
 mp.get_files('data',ext='.jpg')
 | select(lambda x: cv2.imread(x))
 | where(lambda x: x.shape[0]>=1024)
 | select(augment)
 | select(resize)
 | as_list )
\end{minted}

Here we get the list of the image data as a result, but we have lost all the information about the original filenames, and possibly classes of images (if we are solving classification problem). Also, if we want the neural network classifier to use any image metadata, there is no elegant way to pass it along. We could, theoretically, use tuples to group several pieces of data together, but in this case all \texttt{select} functions would look rather ugly, because they would have to operate only on one piece of info inside the tuple, while keeping the rest. This brings us to the idea of passing dictionaries through the pipeline, instead of atomic values.

\subsection{Dictionary generators}

In \mPyPl, we almost always create pipelines that operate on named dictionaries, thus several values can be passed through the pipeline at the same time. We can think of those named fields as 'columns' (if we look at data processing from Pandas perspective), or as object attributes.

For example, suppose we have a directory with image files, and we want to imprint the date of each file on top of the image. In this case, it would be useful to construct a data stream that contains both date/time info, as well as the actual image content. We can get the stream of file names using \verb|get_files| function, and then create a combined stream using the following code:

\begin{minted}{python}
images = (
 mp.get_files('images',ext='.jpg')
 | mp.as_field('filename')
 | mp.apply('filename','image', imread)
 | mp.apply('filename','date', 
     lambda x:str(os.stat(x).st_mtime))
 | mp.apply(['image','date'],'result',
     lambda x: imprint(x[0],x[1]))
 | mp.select_field('result')
 | mp.as_list)
\end{minted}
    
Here, the function \verb|as_field| convert the stream of filenames to named multi-field stream of dictionaries. \verb|apply| takes one or more fields from the stream and computes another field using provided function or lambda-expression. The function \verb|select_field| converts named multi-field datastream back to atomic values, taking only one field\footnote{\texttt{select\_field} can also extract multiple fields, in which case it returns a generator of tuples.} from the stream, which results in the generator of image arrays. At the end we convert this generator to list to store it in memory for future processing \footnote{We could have also converted it to large NumPy array using \texttt{as\_npy}.}

\subsection{\texttt{mdict} and lazy evaluations}

Standard Python dictionaries could have been used, but to allow more flexibility in terms of field computational strategies we introduce special Python class called \texttt{mdict}. It is essentially a dictionary, but each field has an associated computational strategy:

\begin{itemize}
    \item \textbf{Value}, which corresponds to eager strategy, and means that the actual value is stored in the dictionary field
    \item \textbf{LazyMemoized}, which corresponds to lazy evaluation. Dictionary field contains a function to be evaluated upon first request, and then obtained value is stored
    \item \textbf{OnDemand}. An evaluation function is stored in the field, and it is called every time we need to obtain a value
\end{itemize}

We can convert a stream of values into stream of named fields using \verb|as_field| method. A better version of an example described in section \ref{sec_pipe} can be expressed as:

\begin{minted}{python}
images = (
 mp.get_files('data',ext='.jpg')
 | mp.as_field('filename')
 | mp.apply('filename','image', imread)
 | mp.filter('image',lambda x: x.shape[0]>=1024)
 | mp.apply('image','aug_image',augment,
      eval_strategy=mp.OnDemand)
 | mp.apply('aug_image','output',resize,
      eval_strategy=mp.OnDemand)
 | mp.as_list)
\end{minted}

The most often used function in \mPyPl is called \texttt{apply}. It takes names of input and output fields, and a function, which is applied to the input field, and the result is stored in the output field. Thus, \verb|apply| can be considered an analogue of a computed field in Pandas.

Note that in our example we also specify evaluation strategy for the last two fields as \verb|OnDemand| --- this is important for data augmentation to work properly. Eventually during training we would probably loop over data infinitely, and it would cause the \verb|output| field to be re-computed each time we need to get the value, thus each time a new augmentation would be applied.

Sometimes we might need to use \verb|apply| with several input fields -- in this case, we can pass a list of field names as a first parameter, and the corresponding processing function would expect a list of arguments. For example, we can use the following construction to print a text value over an image (suppose \verb|imprint| function expects text as first argument and image as the second):

\begin{minted}{python}
mp.apply(['text','image'],'out',
  lambda x: imprint(x[0],x[1]))
\end{minted}

In addition to \verb|apply|, there are many more useful functions such as \verb|filter|, \verb|fold|, \verb|scan| (cumulative fold), etc. We will not describe all those functions in detail here, and refer you to the documentation available at \url{http://shwars.github.io/mPyPl}.

\subsection{Data Sources}

So far we have only seen \verb|get_files| function that returns a stream of file names from a given directory. There are some more generator functions that can obtain data from common data sources:

\begin{itemize}
    \item \verb|get_xmlstream_from_dir| reads a series of XML files from a given directory and returns their content as named \verb|mdict|'s. It has many parameters to skip or flatten certain fields. One special case of this function is \verb|get_pascal_annotations|, which reads PASCAL VOC annotations for object detection task (which are essentially XML files)
    \item \verb|csvsource| reads a CSV file, and returns corresponding stream of \verb|mdict|'s. Schema is derived from first line in CSV file
    \item \verb|jsonstream| reads a JSON file, which should be a list of dictionaries, and returns corresponding \verb|mdict| stream
    \item \verb|videosource| opens a video file, and returns its frames as a sequence (optionally doing some resizing). Please note that for historic reasons it returns frames as np-arrays, so you need to call \verb|as_field| to convert them to proper named \mPyPl stream
    \item \verb|get_datastream| is the function most commonly used for classification tasks
\end{itemize}

Please not that any Python generator can be used as a data source in \mPyPl.

\subsection{Classification Tasks}

Let us consider the latter \verb|get_datastream| function in more detail, since it is the basis of doing classification with \mPyPl. It expects a base directory, which contains multiple subdirectories --- one per class. It automatically considers those subdirectories to be class names, numbers the classes from 0, and returns \verb|mdict|'s with the fields \verb|filename|, \verb|class_no| and \verb|class_name|. If you want to provide the list of classes and their mapping to \verb|class_no| values, you can also pass a dictionary with \verb|class_no|s and \verb|class_name|s as \verb|classes| parameter.

In machine learning, we often need to split the data into train-test or train-validation-test sets. In \mPyPl, it is normally done by setting a field called \verb|split|. For doing this split, a very flexible \verb|datasplit| function can be used, that can either perform a random split, or save/load split to a file, to make sure we are dealing with the same split each time. Also, sometimes we may want to do split based on some filename pattern, which can be achieved by using \verb|datasplit_by_pattern|.

We also often need to stratify dataset, i.e. make sure there is the same number of samples of each class. This can be done using \verb|stratify_sample|, or \verb|stratify_sample_tt| (the latter also does stratification with respect to train/test split). You can inspect the number of samples for each class using \verb|summary| method.

\subsection{Complete Cats vs. Dogs Example}

Let us consider using \mPyPl to do complete classification task on images --- a classical cats vs. dogs problem described in \citep{Chollet2016}. We will place images of cats in \verb|data/cats| directory, and dogs --- into \verb|data/dogs|. Loading images, resizing them, doing data augmentation and performing train-test split can be done using one pipeline as follows:

\begin{minted}{python}
import mPyPl as mp
import mPyPl.utils.image as mpui
train, test = (
 mp.get_datastream('data',ext=".jpg")
| mp.datasplit(split_value=0.2)
| mp.stratify_sample_tt()
| mp.summary()
| mp.apply('filename','orig_image',imread)
| mp.apply('orig_image','transformed_image', 
   transform.random_transform, 
   eval_strategy=mp.EvalStrategies.OnDemand)
| mp.apply('transformed_image','scaled_image', 
   lambda x: mpui.im_resize_pad(x,size=(150,150)), 
   eval_strategy=mp.EvalStrategies.OnDemand)
| mp.apply('scaled_image', 'image',lambda x:x/255., 
   eval_strategy=mp.EvalStrategies.OnDemand)
| mp.make_train_test_split)
\end{minted}

For data augmentation we use \verb|ImageDataGenerator| from Keras, which can be defined as follows:

\begin{minted}{python}
from keras.preprocessing.image 
import ImageDataGenerator
transform = ImageDataGenerator(
        rotation_range=40, width_shift_range=0.2,
        height_shift_range=0.2,
        shear_range=0.2, zoom_range=0.2,
        horizontal_flip=True, fill_mode='nearest')
\end{minted}

We will omit defining the actual model architecture here, but once it is defined, you can train the model using the following code:
\begin{minted}{python}
args = { feature_field_name: 'image', 
         label_field_name: 'class_id', 
         batch_size: 128 }
hist=model.fit_generator(
      train | mp.infshuffle | mp.as_batch(*args), 
      validation_data=
      test | mp.infshuffle | mp.as_batch(*args))
\end{minted}

\section{\mPyPl and Related Work}

Now that we have introduced the overall architecture of \mPyPl, let us discuss how it relates to other data processing approaches that exist. Comparison of \mPyPl to related projects is summarized in Table~\ref{table_compare}.

\subsection{\mPyPl and Apache Spark}

The most obvious candidate for comparison with \mPyPl is \textbf{Apache Spark}, which also organizes data processing into functional lazy pipelines. However, the main focus of Spark is to power distributed computations on clusters, i.e. Spark pipelines operate on Resilient Distributed Datasets, and are designed to be effectively distributed. On the other hand, the main focus of \mPyPl is to power local computations, for which Spark would be an overkill. 

\begin{table*}[t]
\vskip 0.05in
\begin{center}
\begin{small}
\begin{tabular}{|p{2cm}|p{2cm}|p{2cm}|p{2cm}|p{2cm}|p{2cm}|p{2cm}|}
\toprule
 & \mPyPl & \textbf{Pandas} & \textbf{Apache Spark} & \textbf{ML.NET} & \textbf{Azure ML Data Prep} & \textbf{Azure ML Pipelines}\cr
\midrule
Abstraction & Multi-Field Datastream & Computed Column & RDD & IDataView & DataFlow & Pipeline Step\cr\hline
Lazy & $+$ & $-$ & $+$ & $+$ & $+$ & $-$ \cr\hline
Cluster-ready & $\pm$ & $-$ & $+$ & $-$ & $+$ & $+$ \cr\hline
Succinct syntax & $+$ & $+$ & $\pm$ & $-$ & $-$ & $-$ \cr\hline
Main Focus & single-node large data deep learning tasks 
           & single-node in-memory tables 
           & multi-node cluster computations 
           & .NET ML projects
           & pipelines that can run remotely 
           & MLOps steps \cr
\bottomrule
\end{tabular}
\end{small}
\end{center}
\caption{Comparison of \mPyPl to other data processing engines.}
\label{table_compare}
\vskip 0.1in
\end{table*}

In addition to that, the main advantage of \mPyPl is to provide multi-field named pipelines, which allows concentrating all features within one pipeline in a clear and concise way. This idea, while can probably be implemented on top of Spark\footnote{Such an implementation would be quite difficult due to typed nature of Scala, the language behind Apache Spark. In untyped dynamic language like Python, where untyped dictionaries are first-class citizens, such an implementation is definitely more straightforward.}, is not provided out-of-the-box.

While Spark is a distributed data processing engine, its usage for deep learning has been mostly limited to embarrassingly parallel tasks, such as applying deep learning models to data at scale, or doing hyperparameter optimizations \citep{Databricks2019dlp}. While distributed training is somehow supported, it is mostly through \textbf{Horovod} \citep{Databricks2019ht} third-party library. The same approach of using Horovod for distributed training can be equally well applied to multi-node \mPyPl-powered cluster distributed through Azure Batch (see \ref{section_future} for more information on our plans in this area).

\subsection{\mPyPl and Pandas}

We have already mentioned Pandas here, and noted the similarity between \verb|apply| function and the notion of computed column in Pandas. Pandas represents data in the form of dataframes that are stored in memory and indexed, which allows performing many powerful operations on them, such as pivoting, grouping, flexible masking, etc. However, for deep learning data preparation tasks, many of those operations are an overkill, and mostly linear data processing followed by batching is required. This is exactly the focus of \mPyPl, which provides enough flexibility for deep learning, while allowing data to be processed in batches in a lazy manner, without loading everything into memory.

\subsection{Other Data Pipeline Concepts}

The notion of data processing pipelines exists also in ML.NET in the form of \verb|IDataView| interface, in Scikit.learn, as well as in Azure ML DataPrep SDK. While different approaches may have some advantages over \mPyPl in terms of pipeline execution, they are lacking two main features: ability to define pipeline with multiple named fields, and succinctness of code, achieved through pipe operator overloading. 

Most of the mentioned pipeline approaches are closely related to underlying implementation of the ML engine, be that ML.NET or Azure ML. For example, \verb|IDataView| interface has quite complex architecture, which is required by overall ML platform. The same is true for Azure ML DataPrep SDK, which allows to define pipelines for remote execution on distributed compute. On the contrary, \mPyPl is a very flexible, simple and generic Python tool, which can be used with multiple deep learning and ML frameworks, such as Keras, TensorFlow, Scikit learn, etc. We believe that for many tasks it provides a good compromise between available data processing features and complexity, while resulting in very clean, succinct and functional code. While it is mostly targeting local scenarious, it helps create distributed processing for massively parallel tasks, as we will mention below.

\section{Why \mPyPl Stands for Monadic}

A special note should be made on the name of the library, and why we called it \textbf{monadic}. Strictly speaking, operations on named multi-field data streams do not directly correspond to the notion of monad, as it is known in functional programming \citep{Wadler92}. However, there are some notable similarities, which we believe are enough to justify the name.

In functional programming, monads of type $T$ are represented as some monadic type $M_T$, with two operations defined: $\mathbf{return}: T\to M_T$ and \textbf{bind}: $>>= : M_a \to (f \to M_b) \to M_b$. \textbf{return} simply encapsulates 'normal' type into monadic type, and bind applies a given function $f$ (from 'normal' type to monadic type) to a monadic type, producing the result as a monadic type.

In our case, we are dealing with data streams of dictionaries that can encapsulate many 'named' values at once, thus we would need to alter the \textbf{return} and \textbf{bind} operations to add field name. To do so, let us introduce some notation.

We will denote by $\str{T}$ a data stream of type $T$. By $\md{u}$ we will denote an \verb|mdict| with field name $u$, and by $\mdetc{u}$ we will denote an \verb|mdict| that contains field $u$ and (possibly) some other fields. Just writing $\mathcal{M}$ would mean any \verb|mdict|, no matter which fields it contains. 

We will also use $\mathcal{M}\{u\leftarrow x\}$ to denote \verb|mdict| value with field $u$ equal to $x$, and we will sometimes use $|$ operator as reverse-order function application, i.e. $x | f \equiv f(x)$. We will employ bracket notation for \verb|mdict| indexing, so that for example $\mathcal{M}\{u\leftarrow x\}[u] \equiv x$.

Given those definitions, we can define the following two main functions:
\begin{itemize}
    \item $\mathbf{as\_field}_u : \str{T} \to \str{\md{u:T}}$, which is analogous to \kw{return}
    \item $>>=_u : \str{\mdetc{u:T}} \to (T\to\mdetc{v}) \to \str{\mdetc{u,v}}$, which is analogous to \kw{bind}
\end{itemize}

Here by $u, v, \dots$ we denote field names, and $u:T$ means that field name $u$ is assumed to contain values of type $T$.

Note that \textbf{bind} function $>>=_f$ is not directly used in \mPyPl --- even though it is defined in the library for internal usage and called \verb|__fnapply|. Its direct usage outside the core \mPyPl is discouraged, because it requires some understanding of library internals. Instead, \verb|apply| should be used wherever possible, which is somehow similar in functionality, i.e.
$$\str{\md{u}} | \textbf{apply}_{u,v} f \equiv \str{\mdetc{u}}>>=_u f'$$
where $f'(x) = \mathcal{M}\{u\leftarrow x, v\leftarrow f(x)\}$.

For brevity, we will also denote $\str{\mathcal{M}} | \textbf{apply}_{u,v} f$ as $\str{\mathcal{M}} |>=_{u,v} f$.

For those functions, we can observe that the following laws similar to monadic laws are true:
\begin{itemize}
\item $\mathbf{as\_field}_u \str{T} >>=_u f \equiv \mathbf{map} f \str{T}$ -- left identity. Here, $f : T \to \mathcal{M}$.
\item $\str{\mdetc{u}} >>=_u \mathbf{as\_field}_u \equiv \str{\mdetc{u}}$ -- right identity
\item $\str{\mdetc{u}} >>=_{u} f_v >>=_v g_w \equiv \str{\mdetc{u}} >>=_u (\lambda u . f_v[v] \circ g_w )$ -- associativity. Here we denote by $f_v$ and $g_w$ functions that produce \verb|mdict|s with fields $v$ and $w$, i.e. $f_v : T\to\md{v}$, $g_w : T'\to\md{w}$. 
\end{itemize}

Associativity property does not look very nice, because $>>=$ operator that operates on streams of \verb|mdict|s translates to normal composition $\circ$ when being down-lifted to the function level. This property can be reformulated in terms of \textbf{apply} in the following way:

$$
\str{\mdetc{u}} |>=_u f_v |>=_v g_w \approxeq \str{\mdetc{u}} |>=_u (f_v \circ g_w)
$$

The equivalence here is approximate ($\approxeq$), because in the RHS there field named $v$ is missing.

The latter property turns out to be very useful in practical computations for simplifying the code, and it can be re-stated saying that the two following pieces of code are almost identical:
\begin{minted}[fontsize=\footnotesize]{python}
  (source ...
   | mp.apply(u,v,f)
   | mp.apply(v,w,g)
   ...)
\end{minted}

and

\begin{minted}[fontsize=\footnotesize]{python}
    (source ...
     | mp.apply(u,w,compose(f,g))
     ...)
\end{minted}

\section{Case Study}

Initially, creation of \mPyPl was motivated by one project on complex event detection in videos, which we will briefly describe here. The goal of the project was to be able to detect events in Formula-E racing videos, such as car collisions. Source code of the project, together with more detailed description, can be found in \citep{Valieva2019}.

Among many methods used for event detection in videos (a good review can be found in \citep{Kang16}), there are methods that involve considering two streams of data \citep{Simonyan2014} --- static frame images, as well as dynamic properties of image change over time. Thus, resulting neural network would have quite complex architecture, taking several streams of data as input.

\subsection{Network Architecture and Training}

For the experiment, we have collected a dataset of 5-second video clips with and without collisions from publicly available sources. The dataset contained around 600 positive and negative samples. The lack of samples was mostly because collision events happen to be quite rare during the actual race. 

Due to the relatively small number of samples, we were focusing on manually engineering some features that would allow us to minimize possible overfitting, and make the number of parameters smaller. Minimizing overfitting also means getting rid of the original image pixels, and switching either to CNN embedding domain, or to motion domain represented by either dense, or focused optical flow \citep{Bruhn2005}.

\begin{figure*}
    \centering
    \includegraphics[width=5in]{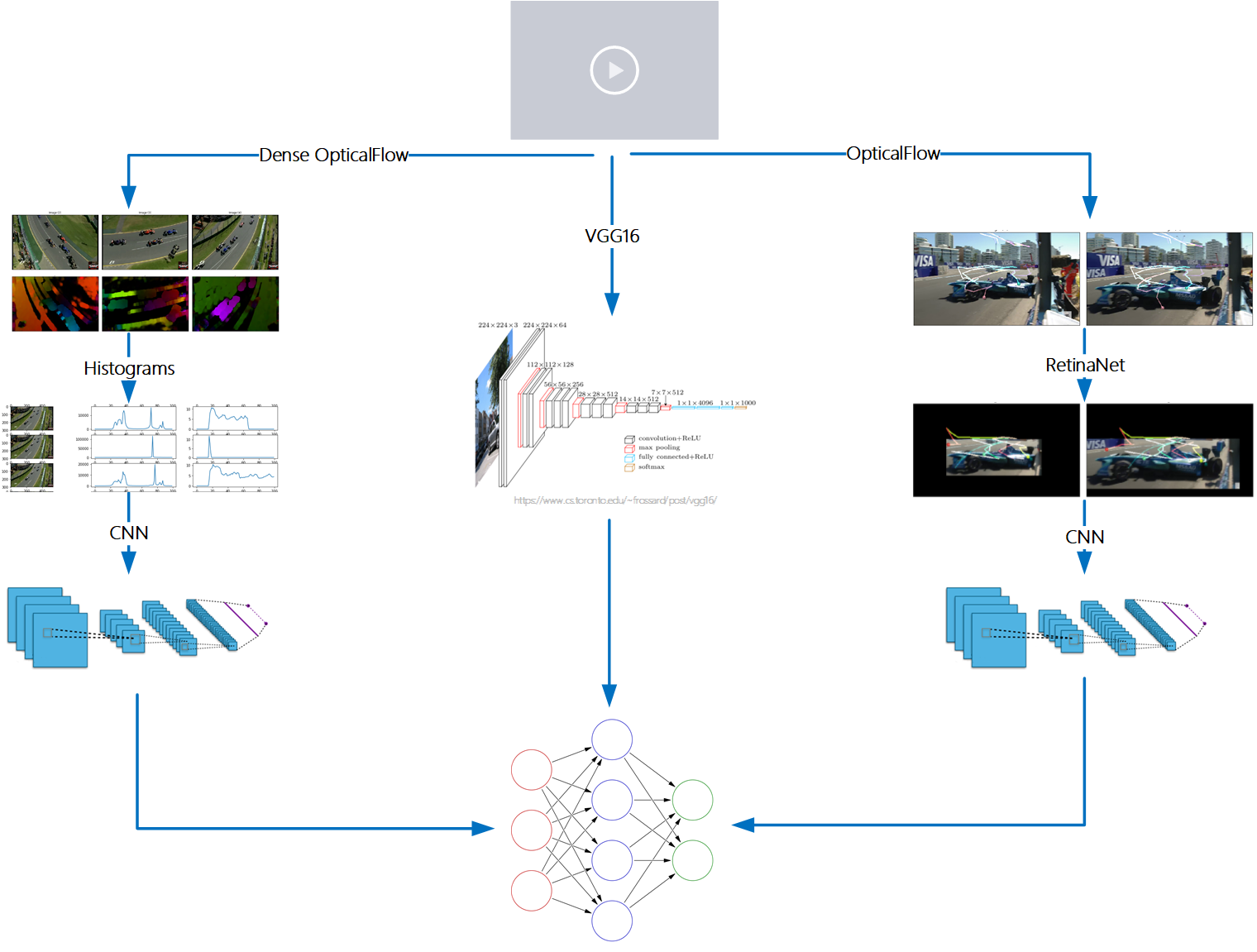}
    \caption{Three-stream pipeline for complex event detection}
    \label{MLWorkflow}
\end{figure*}
    
In our case, we have considered three main data streams (as shown in Fig.~\ref{MLWorkflow}):
\begin{itemize}
    \item Original frames are processed using \textbf{VGG-16 embeddings}, and the resulting vectors are stacked together into rectangular matrix. This matrix is then fed into 2D or 3D CNN network for pattern detection in the embedding space.
    \item \textbf{Dense Optical Flow} field is computed between each pair of frames, the result is converted to polar coordinates and histograms of magnitudes and directions are computed (we used 200 bins). Those histograms between each two frames are stored in $125\times200\times2$ tensor, which is then processed using CNN.
    \item \textbf{Focused Optical Flow} (see Fig.~\ref{FocusedFlow}), which is computed after car detection using RetinaNet\citep{Lin2017}, by tracking features inside detected rectangle. Focused flow vectors are than transformed in the same manner as dense flow.
\end{itemize}

\begin{figure}
    \centering
    \includegraphics[width=3in]{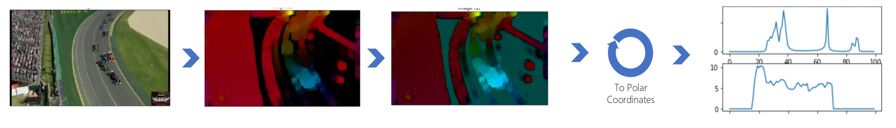}
    \caption{Dense Optical Flow Featurization}
    \label{DenseFlow}
\end{figure}

\mPyPl allows us to create a single data pipeline containing all three data streams, which are than passed onto the neural network. While you can see the complete code in \citep{Valieva2019}, the simplified code below gives the main idea:

\begin{figure}
    \centering
    \includegraphics[width=3in]{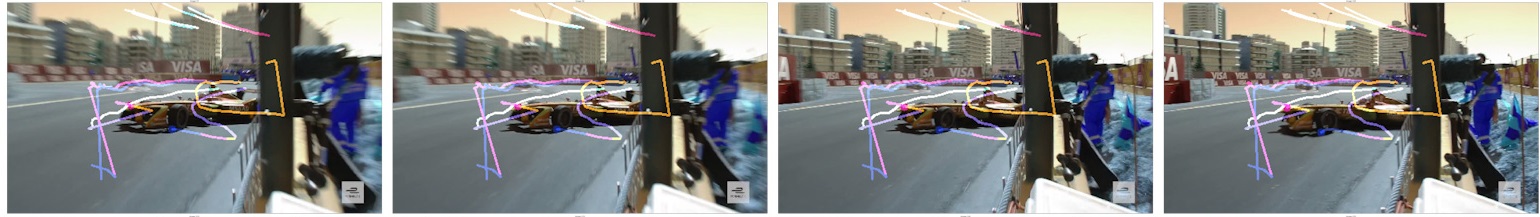}
    \caption{Focused Optical Flow superimposed on original image}
    \label{FocusedFlow}
\end{figure}

\begin{minted}[fontsize=\tinyfont]{python}
trainstream, valstream = (
mp.get_datastream(data_dir)
# Load video
| mp.apply('filename','video',load_video)
# VGG16 embeddings pipeline
| mp.apply('filename', 'raw_vgg', compute_vgg)
| mp.apply('raw_vgg','vgg', zero_pad)
| mp.delfield('raw_vgg')
# DenseFlow pipeline
| mp.apply('video','raw_dflows', compute_dflow) 
| mp.apply('raw_dflows','prep_dflows', zero_pad) 
| mp.apply('prep_dflows', 'res_dflows', 
                        normalize_histograms) 
| mp.delfield(['raw_dflows', 'prep_dflows'])
# RetinaFlow pipeline
| mp.apply('video', 'raw_rflows', compute_optical)
| mp.apply('raw_rflows', 'gradients', calc_gradients)
| mp.apply('gradients', 'polar', to_polar)
| mp.apply('polar', 'histograms', video_to_hist)
| mp.apply('histograms', 'res_rflows', zero_pad_hist)
| mp.delfield(['video','raw_rflows', 
      'gradients', 'polar', 'histograms'])
# Split
| mp.make_train_test_split 
)
\end{minted}

Obtained train and test streams can then be passed onto the network training function:

\begin{minted}{python}
flds = ['res_rflows', 'res_vgg', 'res_dflows']
history = model.fit_generator(
    trainstream | mp.infshuffle 
    | mp.as_batch(flds,'class_id',batchsize=32),
    validation_data=valstream | mp.infshuffle() 
    | mp.as_batch(flds,'class_id',batchsize=32), 
...)
\end{minted}

We want to emphasize the following:
\begin{itemize}
    \item We use special function \verb|delfield| here in order to delete certain fields from the stream. Without those functions, loaded videos would stay in memory for the whole period of training, possibly resulting in out-of-memory conditions. Using \verb|delfield| results in the fact that data stays in memory only during the actual computations, i.e. during minibatch preparation.
    \item If we want to store intermediate results on disk, so that computations can be resumed or performed more quickly, we can use \verb|apply_npy| function, which is the same as \verb|apply|, but stores the intermediate computation results on disk. In case those files are available, they will be loaded from disk without re-computation. This turns out to be very useful for optical flow computations.
    \item In the code sample above, all features for all videos would be computed before the actual training takes place, due to the nature of eager evaluation. If we want the training to start immediately, loading videos from disk and computing features only for each minibatch, we can specify all computations performed in \verb|apply| as \verb|LazyMemoized|. In the current design on the library, all fields that depend on lazy fields should also be declared lazy. Also, if lazy evaluation strategy is used, we should not use \verb|delfield|s that make original fields unavailable, because lazy computations depend on them.
\end{itemize}

\subsection{Video Inference}

To visually inspect relevance of the model, we wanted to produce a clip that includes the collision indicator on top of video, showing the probability of collision predicted by the model. To do so, we need to take 5 second sliding window from the video, pre-compute all features, call trained model to get the result, and then draw the indicator on top of the current frame. This process can also be handled using \mPyPl data streams:

\begin{minted}[fontsize=\tinyfont]{python}
(
  mp.videosource('input.mp4',video_size=video_size)
 | mp.as_field('frame')
 | mp.apply_batch('frame','vggx',get_vgg,batch_size=16)
 | mp.delay('frame','prev_frame')
 | mp.apply(['frame','prev_frame'],'optflow',
                              compute_optflow)
 ...
 | mp.sliding_window_npy(['frame','vgg',
                'optflow','focflow'],size=126)
 | mp.apply('frame','midframe',lambda x: x[60])
 | mp.apply(['vgg', 'optflow', 'focflow'], predict)
 | mp.apply(['midframe','score'],'fframe',imprint)
 | mp.select_field('fframe')
 | mp.collect_video('output.mp4')
)
\end{minted}

\verb|videosource| function opens the video file using OpenCV and produces stream of frames (a generator). We then compute all the features in the same way as we did during training. Some of the code is omitted here for simplicity. Please note the usage of \verb|delay| function to insert previous frame in the pipeline, allowing us to compute pairwise optical flow between frames.

To switch from frame sequence to the sequence of 5-second video intervals we use \verb|sliding_window_npy|. This function groups together specified number of elements into np-arrays of corresponding size, using the sliding window. Thus individual frames and optical flow descriptors are automatically converted into 5-second intervals, and the values are not re-computed for every frame, but are just extracted.

Another thing to be noted here is the usage of \verb|apply_batch|, which is used to feed several values together to the VGG-16 network to optimize computations. It automatically groups together several values, and then un-groups the result into the data stream.

Finally, we call the model on our 5-second interval features, and imprint the result onto the middle frame of the video. All those frames are then grouped together into resulting video using \verb|collect_video| sink function.

\section{Conclusion and Future Work}
\label{section_future}

In this paper, we have presented Python library called \mPyPl, which allows expressing data manipulation tasks in the functional way. We have demonstrated similarity of the proposed architecture to monadic computations, and have presented a real case where usage of \mPyPl simplified the training and inference architectures for video processing quite significantly. We have also compared our work to existing data processing approaches such as Pandas and Apache Spark, and have shown that \mPyPl has its own niche, and it has all the rights to exist.

Our machine learning / data science team in CSE currently uses \mPyPl for almost all of AI/ML related tasks in Python, including object detection, Siamese network training, etc. However, there are further improvements that can be implemented.

During heavy data processing, it can be noticed that not all processor cores are sometimes fully utilized. This is due to the single-threaded nature of Python, and it happens when most of the code is purely Pythonic. To overcome this issue, we are thinking of creating multi-threaded version of \verb|apply| that will spread the computation over available cores.

Another issue is using \mPyPl to parallelize computation on clusters. During data pre-processing, we often need to do embarrassingly parallelizable computations such as optical flow computation or object detection on each clip of the video. This is typically done using Azure Batch service, however, it requires some hand-made data splitting functionality. Using \mPyPl, we can just insert \verb|mp.batch(k,n)| function into the pipeline to separate every $n$-th element with the remainder $k$, i.e. when computing on $n$-node cluster we just need to pass the number of node (from 0 to $n-1$) to the computation script. While this functionality already exists, we are looking for some further way to simplify cluster computations.

We hope that \mPyPl can be useful in a wide range of scenarious outside our internal team, and we would welcome any contributions, issues or comments at the project GitHub repository: \url{http://github.com/shwars/mPyPl}.

\section*{Acknowledgements}

\mPyPl was first developed while working on video event detection project together with Tim Scarfe, whose idea it was to use Pipe library for data pre-processing. We would also like to thank him for using the library extensively in successive projects, and being not only a contributor, but also the most active advocate of it. We would also like to thank other members of our team, who contributed ideas and bug reports, in particular Tess Ferrandez and Claus Matzinger. We would also like to thank our external collaborators, Gianni Rosa Gallina and Clemente Giorio from DeltaTre Innovation Lab for their valuable feedback.

\newpage 

\bibliographystyle{apalike}
\bibliography{biblio}{}












    

\end{document}